\documentclass[12pt]{article}
\usepackage{a4}
\usepackage{graphics}
\usepackage{amsthm}
\usepackage{amstext}
\usepackage{amsmath}
\usepackage{amsbsy}
\usepackage{amssymb}
\usepackage{amsfonts}
\usepackage[vflt]{floatflt}
\usepackage{epsfig}
\def\dm2{\Delta m^2}
\def\sq2{sin^2(2\Theta)}
\def\nubar{\overline {\nu} }

\newcommand{\be}{\begin{equation}}
\newcommand{\ee}{\end{equation}}
\newcommand{\bea}{\begin{eqnarray}}
\newcommand{\eea}{\end{eqnarray}}
\def\NOE{{\em N\raise.5ex\hbox{O \kern-0.47em}E\kern.4em}}

\begin{document}
\title{\bf{Atmospheric neutrinos in a large Liquid Argon Detector}}
\author{G. Battistoni$^a$, A. Ferrari$^{a,b}$, C.~Rubbia$^{b,c}$,
  P.R.~Sala$^a$ and F.~Vissani$^d$}
\maketitle
{\it 
\parindent 70pt
$^a$ INFN Sezione di Milano, Milano, Italy \par
\parindent 70pt
$^b$ CERN, Geneva, Switzerland\par
\parindent 70pt
$^c$ University of Pavia, Italy\par
\parindent 70pt
$^d$ INFN, Laboratori Nazionali del Gran Sasso, Assergi (Aq),
Italy\par}

\abstract{
In view of the evaluation of the physics goals of a large Liquid
Argon TPC, evolving from the ICARUS technology, we have studied 
the possibility of performing precision measurements on atmospheric neutrinos.
For this purpose we have improved existing Monte Carlo neutrino event generators
based on FLUKA and NUX by including the 3--flavor oscillation
formalism and the numerical treatment of Earth matter effects. 
By means of these tools we have studied 
the sensitivity in the measurement of $\theta_{23}$ 
through the accurate measurement of $\nu_e$'s. The updated values for
$\Delta m^2_{23}$ from Super--Kamiokande and the  mixing parameters as obtained
by solar and KamLand experiments have been used as reference input, while
different values of $\theta_{13}$ have been considered.
An exposure larger than 500 kton yr seems necessary in order to achieve
a significant result, provided that the present knowledge of systematic
uncertainties is largely improved.
}
\section{Introduction}
\label{sec1}
In the framework of the design study for a very large Liquid Argon TPC
evolving from the present status of the ICARUS project\cite{icarus},
it is important to start the discussion of the different physics goals that
should be addressed by a multipurpose detector of this kind.
The topic of atmospheric neutrinos is certainly one of the relevant chapters.
Indeed, after the results coming mostly from Super--Kamiokande\cite{skres}, there is
still a scientific interest in continuing the study of atmospheric
neutrinos.  First of all it is necessary to confirm the results from 
SK with a technology capable of a large reduction of experimental
systematics with respect to water \v Cerenkov. In second place, many
authors have pointed out the possibility of exploring subleading
contributions in the oscillation matrix. This would allow, in particular,
a possible precision measurement of $\theta_{23}$ and the discrimination 
of the normal vs. the inverted hierarchy in the neutrino mass scale.
These are tiny effect, and a very large Liquid Argon TPC 
with a sensitive mass of the order of several tens of kilotons 
seems necessary to reach the exposure necessary for these proposed analyses.
The preliminary work here presented aims to investigate quantitatively the
attainable performance of such a large LAr detector in the field of
atmospheric neutrinos

In section \ref{sec2} we describe in more detail
the motivations of physics for this study 
and provide a general discussion of the effects we are 
going to discuss. In section \ref{impl}, we present the
model implemented in our simulation to include
three flavor oscillations and matter effects.
Then, in section \ref{sec4} we discuss the possible results 
with atmospheric neutrinos in the
limit of a very high exposure.

\section{Atmospheric neutrinos and 3--F oscillations\label{sec2}}
A large number of results regarding atmospheric\cite{skres},
solar\cite{solarexp}, reactor\cite{kamland,chooz}  and
accelerator neutrinos can be accounted for assuming that the 
ordinary 3 neutrinos have mass and mix among them \cite{strumia}.
In this framework, flavor oscillations
are fully described by assigning 3 mixing angles
and 1 CP violating phase.
We do not know the type of mass hierarchy, and 
3 of these quantities are still unmeasured:
The size of $\theta_{13}$, the deviation of $\theta_{23}$ 
from maximal mixing, and the CP violating phases $\delta$
(see sect.~\ref{impl} for their definitions).
The first unknown quantity is particularly important, for 
$\theta_{13}$ controls the size of three flavor oscillations.
Although there are little doubts that such a 3$\nu$ 
framework is appropriate to describe these data, 
it is also fair to say that any experimental indication 
that we have is appropriately described by 2 flavor (2F) 
oscillations and 3 flavor (3F) effects have been not yet seen. 

We recall that there 
are several theoretical approaches to understand neutrino 
masses, and possibly to predict unknown quantities 
as $\theta_{13}$. The problem of fermion masses is known to be a difficult one,
and caution is necessary in interpreting any prediction 
or theoretical expectation. 
However, several models suggest that $\theta_{13}$ is 
not undetectably small (see e.g., \cite{so10} for SO(10) models); 
rather, $\theta_{13}$ often happens to be close to the 
experimental limit of about $10^{\circ}$. Similarly, there are several models
where deviations of $\theta_{23}$ from maximal mixing amount to several 
degrees. Presumably, these expectations can be considered in agreement with 
the common sense: in absence of any strong reasons to assume the contrary,
$\theta_{13}$ and $\theta_{23}-45^\circ$ should be not too small.
Certainly, experimental investigations of $\theta_{13}$ and the 
search of 3F effects in oscillations are among the most important 
goals still to be achieved and will have a profound 
impact on what we know about neutrinos.

One of the most important results emerging from the 2F analysis of 
atmospheric neutrino experiments is that the $\theta_{23}$ mixing angle,
which is defined  
in the range 0-$\pi$/2, is found to be compatible with the
full mixing value of $\pi$/4. The measurement of this
mixing angle comes from the up-down asymmetry of muon--like events, and in
particular of the higher energy events, as the Multi--GeV selection in
Super--Kamiokande, where the direction correlation between neutrino and
lepton is strong. Actually, such up/down ratio, which in the average
results very close to 1/2, gives a measurement of $sin^2 2\theta_{23}$ and
therefore it is not possible in this way to distinguish the octant of
$\theta_{23}$. The present result is $sin^2 2\theta_{23}$ = 1 $\pm$ 0.01,
which corresponds to $sin^2 \theta_{23}$ = 0.5 $\pm$ 0.05
In order to achieve a better understanding of the properties of neutrino mixing,
it is of the utmost importance to measure precisely the value of
$\theta_{23}$ and to measure its deviation from the full mixing value.

The possibility of performing a different measurement emerges in the
framework of the 3-flavor oscillation scenario. Beside the main 
part due to $\nu_\mu-\nu_\tau$ oscillations 
certainly we have further small effects of 
$\nu_e$ oscillations with ``solar frequency'' 
and possibly, other $\nu_e$ oscillations 
with ``atmospheric frequency'' due to $\theta_{13}$.
On top of these two effects, essentially of 2F character,
there are also genuine 3F effects 
due to the interference between the various amplitudes of transition.

The following analytical considerations help 
to understand the underlying physics. In general, one can write the formal
solution for the amplitude 
${\cal A}_\nu$ of neutrino propagation as
\begin{equation}
{\cal A}_\nu=\mbox{Texp}[-i\int dt {\cal H}_\nu(t)]=
R_{23}\Delta\cdot 
\left(
\begin{array}{ccc}
a_{11} & a_{12} & a_{13} \\
a_{21} & a_{22} & a_{23} \\
a_{31} & a_{32} & a_{33} 
\end{array}
\right)
\Delta^* R_{23}^t
\label{smart}
\end{equation}
where $R_{23}$ are rotations by the angle $\theta_23$ in the $23$ plane.
The elements $a_{ij}$ form an unitary matrix that depends on 
$\theta_{13}$  and $\theta_{12}$ but not on $\theta_{23}$ 
or on $\delta$ (of course, it depends also on 
$\Delta m^2$'s and on the `kinematical variables' 
$E$, $L$  and $\theta_Z$). From the explicit form of the 
hamiltonian ${\cal H}_\nu$ one easily shows that in the limit $\theta_{13}=0$,
one has $a_{13}=a_{23}=0$, so the effects due to $\theta_{12}$
are described by $a_{12}$ (`$\theta_{12}$-driven' or `solar frequency'
oscillations).
Similarly, in the limit $\theta_{12}=0$,
one has $a_{12}=a_{23}=0$, so the effects due to $\theta_{13}$
are described by $a_{13}$ (`$\theta_{13}$-driven' or  `atmospheric frequency'
oscillations).

In general, one can get
$a_{ij}$ from a numerical calculation. In the case of 
vacuum oscillations, however, the equations can be integrated
and give for instance:
\begin{equation}
\left\{
\begin{array}{ll}
a_{12}=s_{12} c_{12} c_{13} (e^{-i\varphi}-1), & \varphi=\Delta m^2_{12} L/2E \\
a_{13}=s_{13} c_{13} [(e^{-i\Phi}-1)-s_{12}^2 (e^{-i\varphi}-1) ],
& \Phi=\Delta m^2_{13} L/2E 
\end{array}
\right.
\end{equation}
Here we are using the standard notation $s_{ij}$, $c_{ij}$ for
$sin\theta_{ij}$ and $cos\theta_{ij}$ respectively.
It is evident that these amplitudes have properties just discussed in the 
limiting cases $\theta_{12}\to 0$  and $\theta_{13}\to 0$.

From eq.~(\ref{smart}) 
one obtains expressions for 
the probabilities of survival or of conversion,
e.g.:
\begin{equation}
\left\{
\begin{array}{ll}
P_{e\to e}=|({\cal A}_\nu)_{11}^2|=1-|a_{12}^2|-|a_{13}^2| \\
P_{\mu\to e}=|({\cal A}_\nu)_{12}^2|=|c_{23}\ a_{12}\ e^{i\delta} 
\ +\ s_{23}\ a_{13}|^2 
\end{array}
\right.
\label{cinq}
\end{equation}
The interference term in the latter expression is a prime example of 
a genuine 3 flavor (3F) effect, that disappears in the 
limits $\theta_{12}\to 0$ and $\theta_{13}\to 0$. It should be noted 
that the dependence of this interference term on $a_{13}$ is linear,
thus it depends mildly on $\theta_{13}$. (It should be 
recalled that the CHOOZ or reactor limit on $\theta_{13}$, 
instead, scales with $\theta_{13}^2$.)

Using eqs.~(\ref{cinq}) and following the discussion in \cite{peres},
let us introduce the flavor ratio $r=F_\mu^0/F_e^0$.  The variation of the 
electron neutrino flux 
$F_e=P_{e\to e} F_e^0+P_{\mu\to e} F_\mu^0$ 
reads:
\begin{equation}
\frac{F_e}{F_e^0}-1=
(r c_{23}^2 -1) |a_{12}^2| +
(r s_{23}^2 -1) |a_{13}^2| 
+2 r s_{23} c_{23} \mbox{Re}[a_{13}^* a_{12} e^{i\delta}]
\end{equation} 
The first two contribution are small for low energy neutrinos,
since $r\sim 2$ and the $\theta_{23}$ mixing is close to maximal:
$c_{23}\sim s_{23}\sim 1/\sqrt{2}$. It should be noted that these
two terms correspond largely to 2 flavor effects.
In fact, from what told above we know that
in the limit $\theta_{13}=0$, $a_{13}=0$  
thus from the first of eqs.~(\ref{cinq}) 
$|a_{12}^2|=1- P_{e\to e}(\Delta m^2_{12},\theta_{12})$ whereas
in the limit $\theta_{12}=0$, $a_{12}=0$ thus  
$|a_{13}^2|=1- P_{e\to e}(\Delta m^2_{13},\theta_{13})$.
The third term is the only one that it is not suppressed by the flavor ratio
$r\sim 2$. In fact, this is exactly 
the interference term, namely, the genuine  3F effect.
Note finally that the three terms work in 
different ways: the first one increases the electron flux 
if $\theta_{23}<45^\circ$, the second one 
if $\theta_{23}>45^\circ$, the last, can increase
or decrease depending on $\delta$. 
In other words, the ``solar'' sector of the neutrino mixing (described by
the $\Delta m^2_{12}$ and $\theta_{12}$ parameters) influence the
rate of $\nu_e$ events (in particular in the SubGeV
range) with respect to the no--oscillation case, with a
different sign according to the value of $\theta_{23}$, even in case
$\theta_{13}$ is very small or null. The effect vanishes totally instead
for the maximum mixing case $\theta_{23}$ = $\pi$/4.
Since the solar neutrino and KamLand[2, 3] experiments have obtained a
remarkably precise determination of two parameters $\Delta m^2_{12}$ and
$\theta_{12}$, in principle the proposed measurement is possible,
provided that systematics uncertainties are kept under control.

The MSW effect\cite{msw}
(``matter effects'') is relevant 
in specific energy regions, and in particular
this happens where vacuum term is comparable with the matter term.
For instance, the vacuum term at 
$\Delta m^2=10^{-4}$ eV$^2$ and $E=0.5$~GeV is 
$\Delta m^2/2 E\approx 5\cdot 10^{-4}$ eV$^2$/MeV
while the matter term at $\rho_e=2.5$ $e^-$ moles/cc is
$\sqrt{2} G_F N_e=9\cdot 10^{-4}$ eV$^2$/MeV
(as usual, $N_e=N_A\rho_e$). An interesting consequence is that, due to
the MSW effect, $\theta_{13}$ oscillations (driven by 
$\Delta m^2_{atm}\sim 2.5\cdot 10^{-3}$~eV$^2$)
are amplified for $E \sim 2-6$ GeV. This could be of 
particular interest in long-baseline experiments, while, 
for statistical reasons connected to the steeply falling energy spectrum,
atmospheric neutrino experiments are less sensitive in this energy range. 
Therefore in this work we focus on low energy events and neglect measurements 
aimed to measure $\theta_{13}$ and more in general MultiGeV events.
We remark here that one of the most appealing advantage offered by the
Liquid Argon technology with respect to water \v Cerenkov detectors, is the
much larger efficiency in SubGeV investigation, in conjunction to the
partial possibility of a better reconstruction of their kinematics. In fact
an ICARUS--like detector can reconstruct recoiling protons down to a rather
low energy threshold.

There is another important measurement connected with 3F oscillations
including matter effects which can
be in principle performed using atmospheric neutrinos. This is the
discrimination of normal from inverted mass hierarchy (the sign of $\Delta
m^2_{13}$). Large matter effects occur for neutrino if $\Delta m^2_{13}>0$
and for antineutrinos if $\Delta m^2_{13}<0$ and this is again more
significant in the MultiGeV region. As suggested in \cite{petcov05}, a
detector capable of distinguishing on an event by event basis $\nu_\mu$
from $\nubar_\mu$ in the range from about 3 to 10 GeV is ideal for this
purpose. Such a separation capability can be performed by means of a
magnetized detector: there is not yet a convincing design of large Liquid
Argon coupled to a magnet. Furthermore, a low density detector has a low
containment efficiency for MultiGeV muons. 
Therefore we also will not further discuss this suggestion, 
at least in this preliminary work which concerns
the possibilities of further investigation of atmospheric neutrinos with a
large Liquid Argon detector.

\section{The simulation code for 3F oscillations with matter effects\label{impl}}

The discussion of the previous paragraph introduces the need of 
calculating predictions taking into account with the maximum possible
accuracy 3F oscillation coupled with matter effects.
A numerical code has been designed for this purpose and we 
recall here the basic formalism that has been implemented for this
purpose. 

The propagation of neutrinos or antineutrinos in a medium 
with electron density $N_e$ that varies along the neutrino
trajectory (parameterized by the time $t$)
is described by Schr\"{o}dinger-like 
evolution equations\cite{msw}:
\begin{equation}
\left\{
\begin{array}{lll}
{\cal H}_\nu(t)&=& U\cdot
\mbox{diag}(m^2_i)\cdot U^\dagger/(2E) +
\sqrt{2} G_F N_e(t)\; \mbox{diag}(1,0,0) \\
{\cal H}_{\overline{\nu}}(t)&=& U^*\cdot
\mbox{diag}(m^2_i)\cdot U^t/(2E) -
\sqrt{2} G_F N_e(t)\; \mbox{diag}(1,0,0) 
\end{array}
\right.
\end{equation}
where the neutrino mixing matrix 
includes the already mentioned three rotations
through the 23, 13 and 12 planes with angles 
$\theta_{23}$, $\theta_{13}$ and $\theta_{12}$ 
and a CP violating phase $\delta$
\begin{equation}
U=R_{23}\Delta R_{13} \Delta^* R_{12},\ \ \ \Delta=
\mbox{diag}( 1 , 1 ,  e^{i\delta} )
\end{equation}
In the following, we will discuss the case of neutrinos, 
however it is clear that the same discussion applies to
antineutrinos inverting the sign of $N_e$ and of 
the CP-violating phase $\delta$. We use the 
notation $\Delta m^2_{ij}=m^2_j-m^2_i$,
and adopt the convention that  
$\Delta m^2_{12}=\Delta m^2_{sol}>0$,
whereas $\Delta m^2_{13}=\pm \Delta m^2_{atm}$ 
if the mass hierarchy is normal or inverted respectively.
In other terms, we declare that $\nu_3$ 
is respectively the heaviest or the lightest neutrino.

\begin{figure}
\begin{center}
\includegraphics[width=13.5cm]{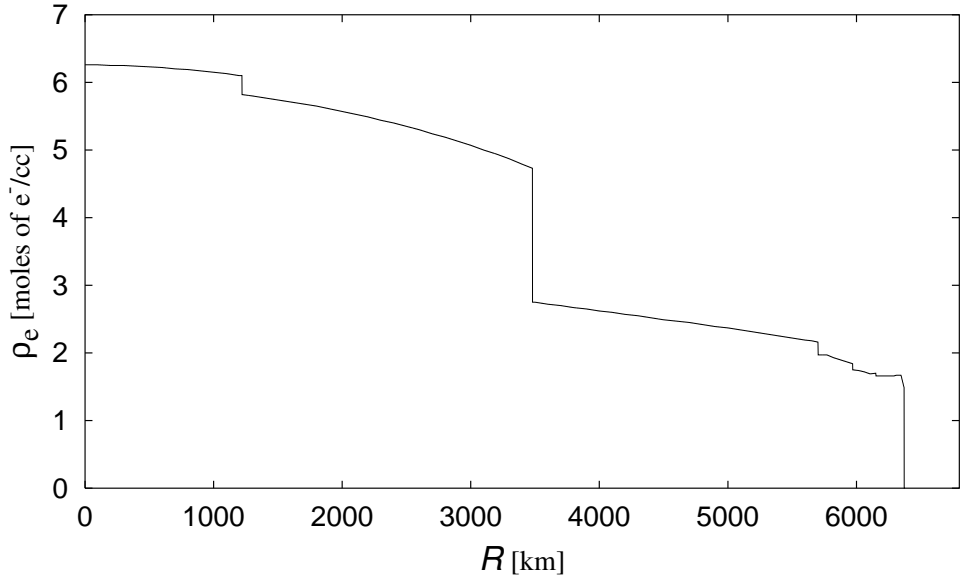}
\caption{Density profile of the Earth according to the PREM
  model\protect\cite{prem}.
\label{figprem}}
\end{center}
\end{figure}

In the code, the evolution equations above are solved 
resorting to numerical methods. 
Being concerned with atmospheric events,
we consider neutrinos of energy $E$ 
that reach a detector 
with zenith angle $\theta_Z$ after a path length $L$;
if $\theta_Z>90^\circ$, these neutrino 
traverse the Earth and therefore meet a non-zero 
electron density $N_e$. We describe the Earth electron density by
adopting the PREM model\cite{prem}. 
which gives the radial profile of the Earth's density
as shown in Fig. {\ref{figprem}}

We solve the resulting evolution equations through 
a straightforward numerical integration.\footnote{We 
use the {\tt ODEINT} subroutine given 
in Numerical Recipes\cite{nrweb}, with the {\tt yscal(i)}
variable set to unity. We obtained a numerical 
precision in the determination of the 
oscillation probabilities much better than one 
part per million for all neutrinos we propagated. 
This tight requirement
is much more than what we actually need, but the required 
computational time is small when compared with the time to
simulate an event.}
Given a neutrino (or antineutrino), we calculate the probabilities that
it will be detected as $\nu_e$, $\nu_\mu$ or $\nu_\tau$ and 
finally, sort the detected flavor according to
these probabilities. 

This code has been coupled to a neutrino interaction generator (for both CC and NC) 
which makes use of the FLUKA\cite{fluka} and NUX\cite{nux} libraries.
Quasi-elastic (QE) and deep inelastic scattering (DIS) are considered. 
The QE part is managed by FLUKA and according to the physics choice of NUX,
the resonance region is considered in average, according to the duality
principle, by the DIS part.  
FLUKA provides the nuclear environment for all reactions. Quantum and
nuclear effects, including reinteractions in the nucleus, are simulated in
detail.
This event generator was  originally developed within the ICANOE
proposal\cite{icanoe} with the possibility of invoking 2-flavor
oscillations (full mixing) in vacuum.
Now the present version has been upgraded so to include the general
3-flavor oscillation code coupled to the above described numerical solution of the
differential equation for the transport of neutrino amplitude through the 
Earth.

Oscillation parameters are provided by means of a data card file
({\em osc.datacards}), where $\Delta m^2$ values must be given
in eV$^2$ and mixing angles in degrees. The CP violating phase $\delta=
(0-360^\circ)$
is also considered, and the user can choose the ``direct'' or ``inverse''
hierarchy. Matter effects can be switched off.
A typical example of the osc.datacard file is reported below:

\begin{verbatim}
#Data cards for atm. neutrino oscillation code
# Theta12 in degrees
THETA12  32.3
# Theta13 in degrees
THETA13   0. 
# Theta23 in degrees
THETA23  45.
# DeltaM12**2 in eV**2
DMSQ12   8.2D-5
# DeltaM23**2 in eV**2
DMSQ23   2.6D-3
# CP violation phase in degrees
DELTA    0.
# Hierarchy: normal: 1, inverted: -1
HIERARCHY 1.
# Activate matter effects: on: 1, off = 0
MATTER 1.
# end of data cards
END
\end{verbatim}

These drivers require as input the neutrino fluxes. For atmospheric
neutrinos we start from the ``FLUKA'' fluxes of 2001\cite{flukanu}.
which are calculated for 3 specific geographic locations. For the purpose of this 
work, we limited ourself to the LNGS case.

As a default choice the interactions are performed against argon nuclei,
but any other nucleus can be considered. The natural isotopic composition
of any given nucleus is automatically provided by the FLUKA database.
These generators produce in output the detailed kinematical information of
the initial, 
intermediate and final state of the interaction and, optionally, the binary
input for a full simulation of detector response as deriving from ICARUS.

\section{Atmospheric neutrinos in a large Liquid Argon TPC
\label{sec4}}

We have simulated a single experiment with 
an high exposure, simulating the statistical sample of CC interactions of
atmospheric neutrinos at Gran Sasso expected in 1000 kton~yr.
We have considered, beyond the unoscillated case, three cases for
$\theta_{23}$ well inside the present limit from atmospheric experiments: 
40$^\circ$, 45$^\circ$ and 50$^\circ$. 
The fit results from the review of ref.\cite{strumia}  have been used as
reference for $\Delta m^2_{23}$, $\Delta m^2_{12}$ and $\theta_{12}$
(2.5 10$^{-3}$ eV$^2$, 8.0 10$^{-5}$ eV$^2$, 34$^\circ$), 
while $\theta_{13}$ has been varied between 0$^\circ$ and 10$^\circ$,
({\it i.e.} about the existing limit from the CHOOZ
experiment\cite{chooz}). For the work described here, the CP violation
phase $\delta$ has been left to 0$^\circ$ and only the direct hierarchy of
masses has been considered. 

\subsection{Distribution in zenith angle and $L/E$\label{sub1}}
\begin{figure}
\begin{center}
\begin{tabular}{c}
\includegraphics[width=10cm]{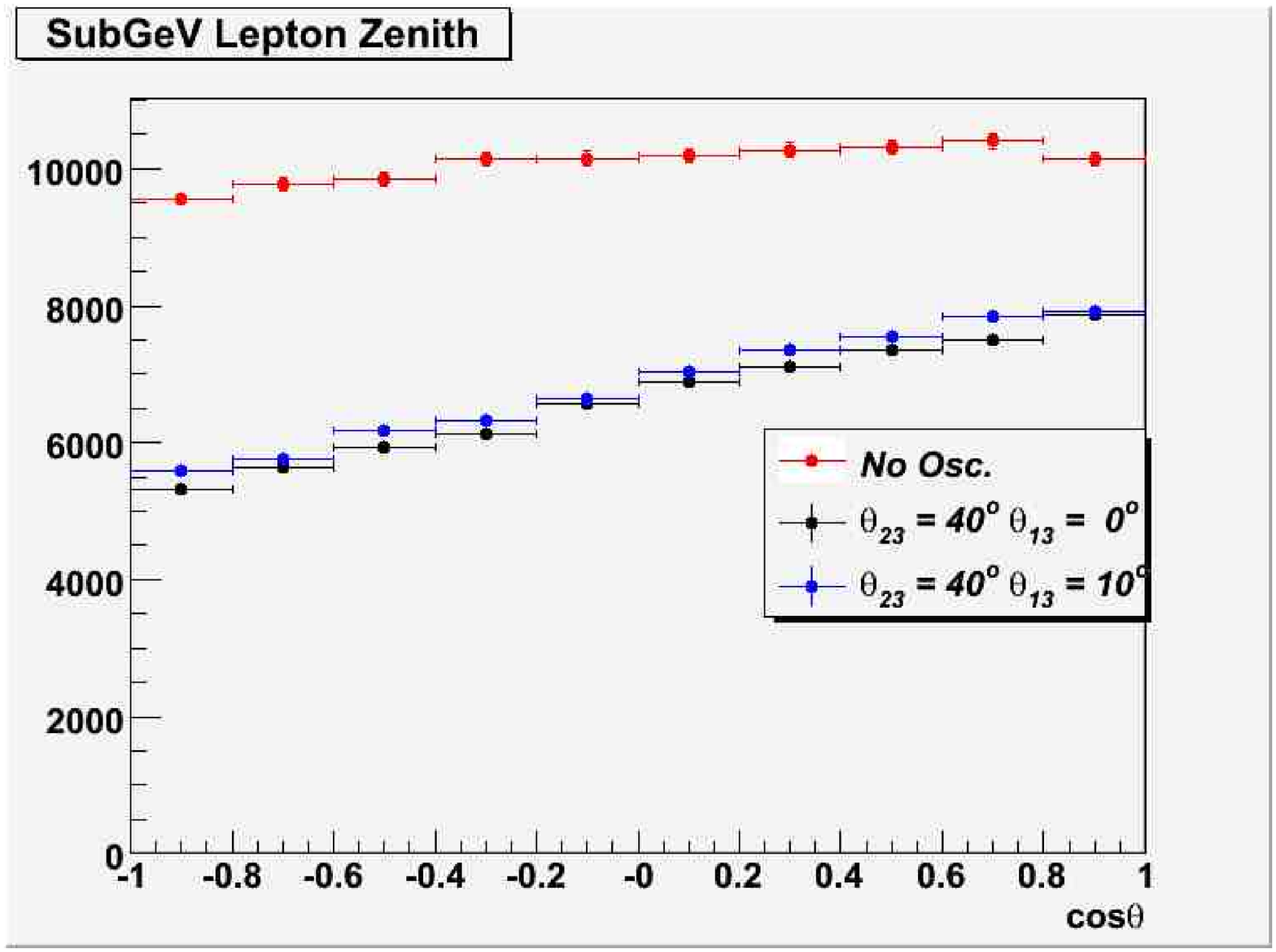} \\
\includegraphics[width=10cm]{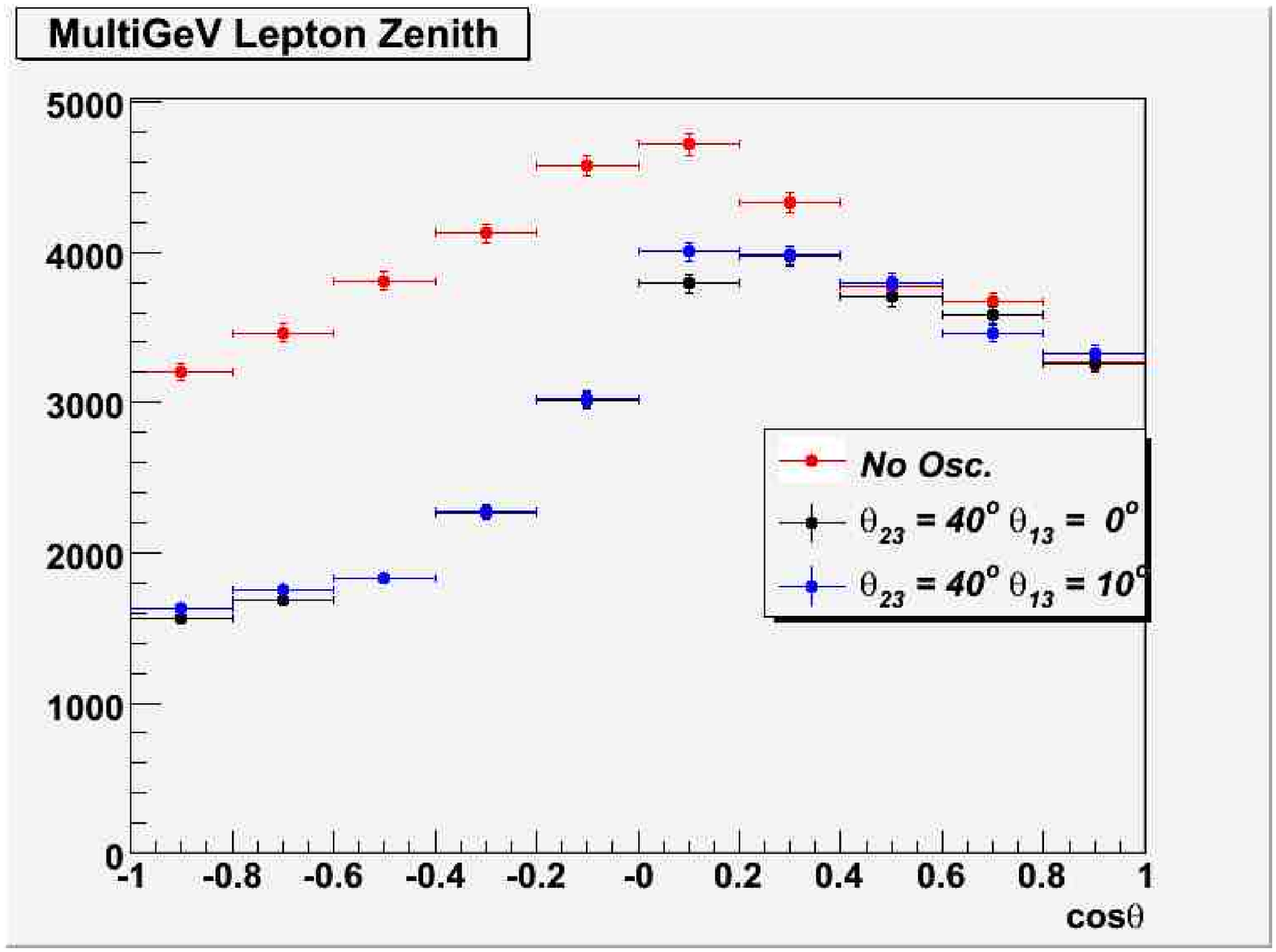} 
\end{tabular}
\caption{Zenith angle distribution for SubGeV (top) and MultiGeV (bottom)
  $\mu$-like events. The non oscillated and
  oscillated predictions for $\theta_{23}$=40$^\circ$ and two different
  values of $\theta_{13}$ are reported. Here only the lepton direction
has been used.\label{nu2}}
\end{center}
\end{figure}

Before entering into the details of the measurement of the octant of
$\theta_{23}$ it is worthwhile considering the ``standard'' measurements
with atmospheric neutrinos. For instance, Fig.\ref{nu2} shows the
expectation for the zenith angular distribution of muon neutrino events, when only
the lepton direction is measured, for SubGeV and MultiGeV regions in the
case of full mixing ($\theta_{23}$ = 45$^\circ$). The two extreme values of
$\theta_{13}$ are considered in the oscillation predictions. 

\begin{figure}
\begin{center}
\includegraphics[width=10cm]{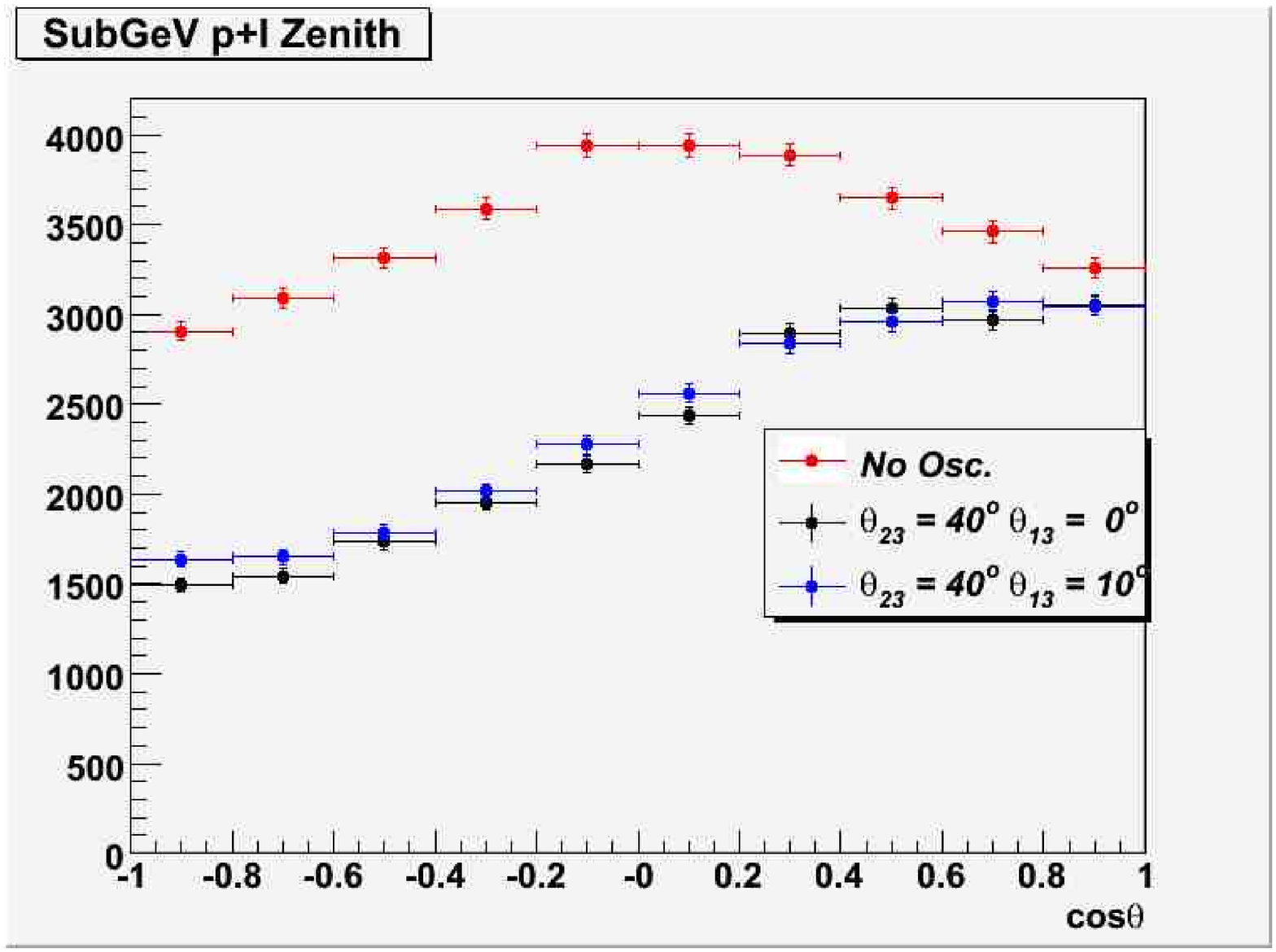}
\caption{
Zenith angle distribution for SubGeV
  $\mu$-like events where both lepton and recoiling proton kinematics has 
been reconstructed (E$_{kin}>$ 50 MeV). The non oscillated and
  oscillated predictions for $\theta_{23}$=40$^\circ$ and two different
  values of $\theta_{13}$ are reported.\label{nu3}}
\end{center}
\end{figure}

This plot already gives a quantitative idea of the difficulty of
measuring $\theta_{13}$ with atmospheric neutrinos.
However, a Liquid Argon TPC gives the chance of performing an improved
measurement of the neutrino direction. This is already possible, in part,
in the SubGeV range. Notwithstanding the intrinsic smearing due to Fermi
motion, a better correlation with original neutrino direction can be
maintained, as shown in Fig.\ref{nu3}, where the kinematic information of
possible recoiling proton has been added, assuming a detection threshold 
of 50 MeV for the kinetic energy of an hadron. The expected enhancement
around the horizontal direction is clearly visible. Despite the reduction
in statistics, the separation between the two extreme cases of
$\theta_{13}$ is slightly improved. 

\begin{figure}
\begin{center}
\begin{tabular}{c}
\includegraphics[width=10cm]{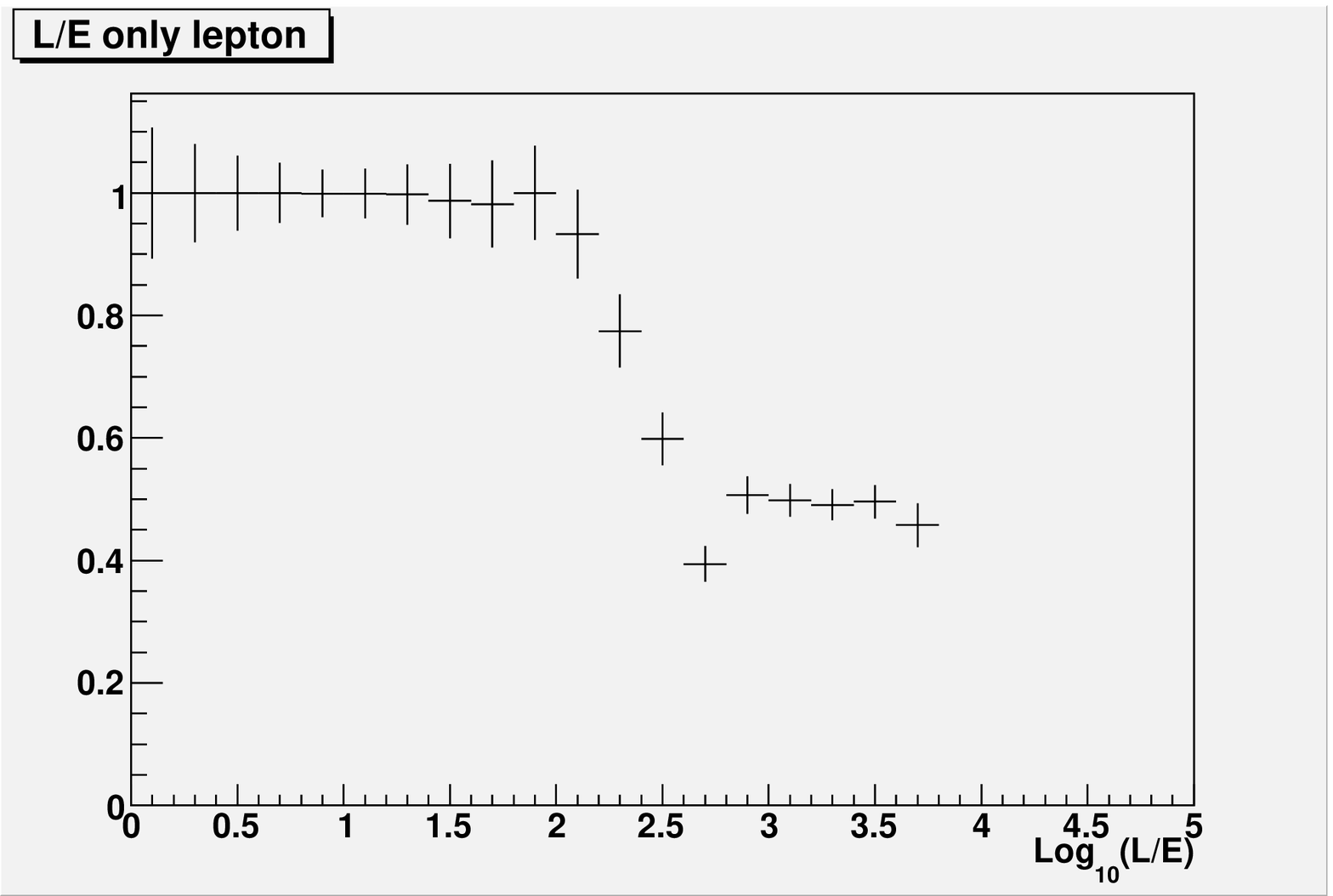} \\
\includegraphics[width=10cm]{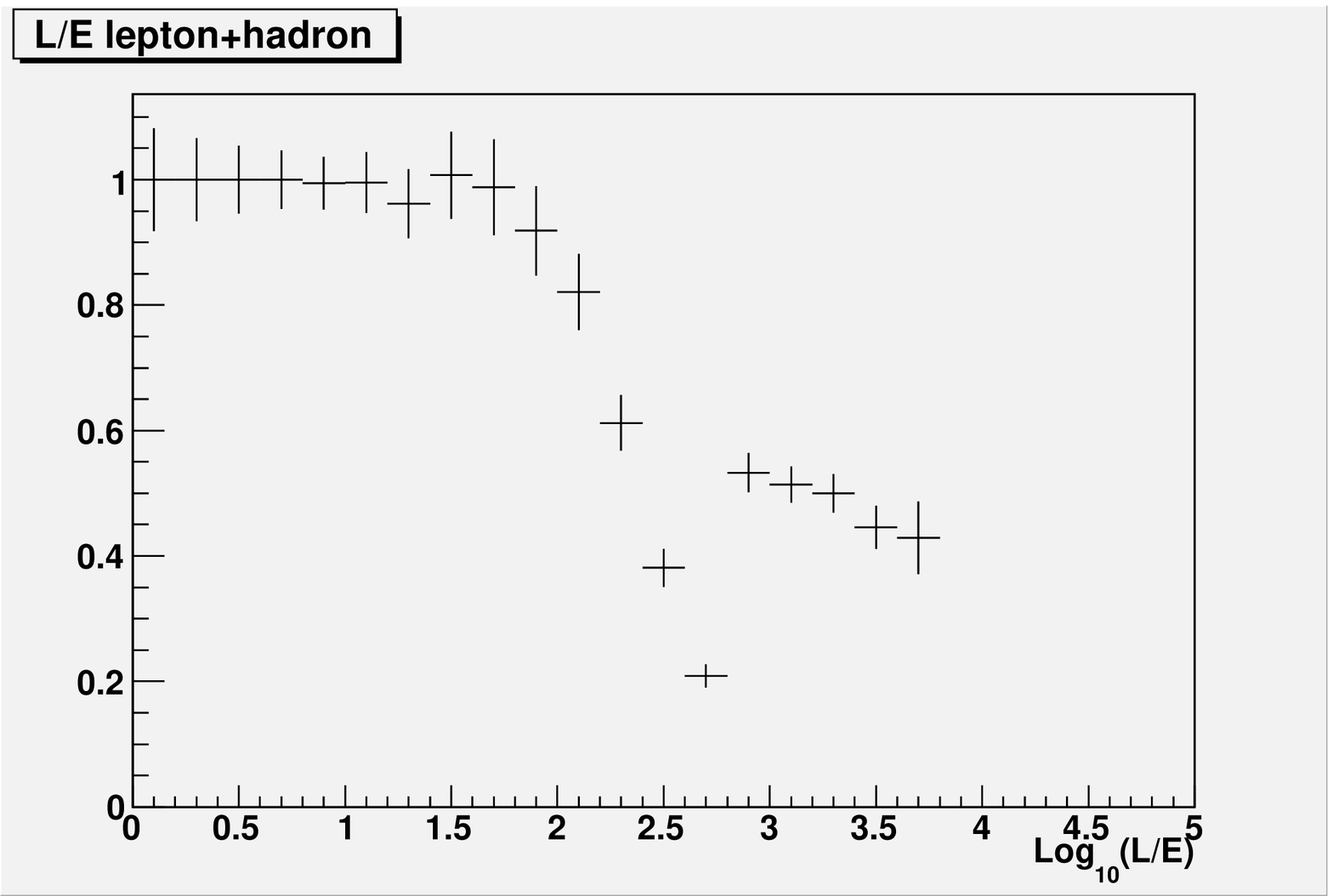} 
\end{tabular}
\caption{Ratio of event rate  with respect to
the non oscillated prediction as a function of $\log{L/E}$.
L is measured in km, while E is in GeV. Top panel: only lepton direction is
reconstructed. Bottom panel: both lepton and charged hadrons are
reconstructed (E$_{kin}>$ 50 MeV). 
\label{nu4}}
\end{center}
\end{figure}

The above considerations suggest that the results from a large Liquid Argon
TPC can be used to perform a significant identification of the oscillation
pattern by means of the plot of the ratio of event rate  with respect to
the non oscillated prediction as a function of $\log{L/E}$. This is
reported in Fig.\ref{nu4} (L is measured in km, while E is in GeV). 
There we have followed the usual method of mirroring the
direction of downward going events to obtain a model independent prediction
for upward going neutrinos. In the upper panel of Fig.\ref{nu4} only the
lepton direction has been considered, while in the bottom panel the
kinematics of the hadronic system (for particles above threshold) has been
included: the significance in pattern recognition is clearly improved.
 
We estimate that, for this measurement, an exposure of 500 kton yr would be
already sufficient to provide a very significant result.

\subsection{Low energy electron events and $\theta_{23}$\label{sub2}}
\begin{figure}
\begin{center}
\includegraphics[width=10cm]{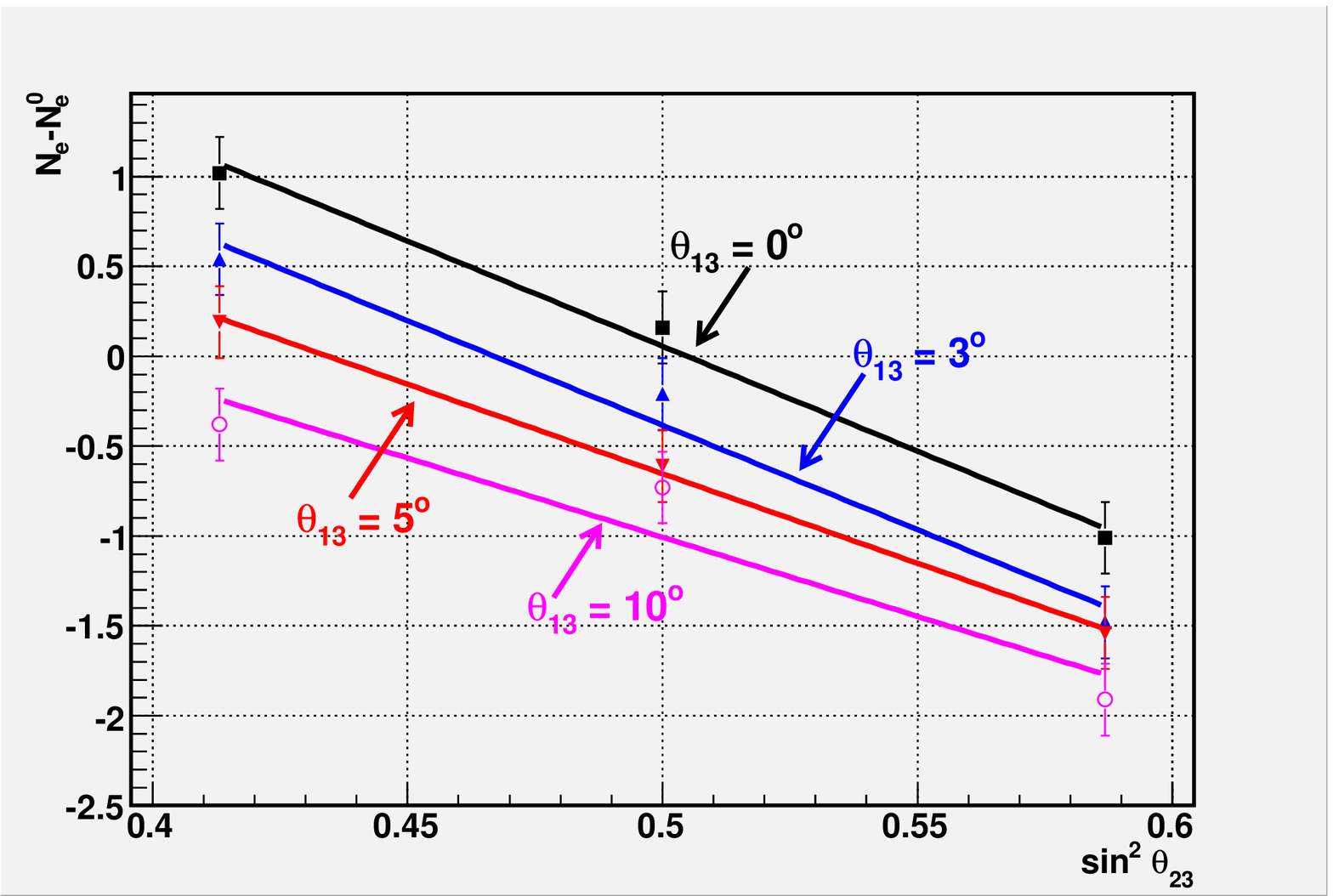}
\caption{Deviation of the SubGeV $e$-like event rate (in events/kton yr)
from the no-oscillation
  prediction as a function of $sin^2\theta_{23}$.\label{nu5}}
\end{center}
\end{figure}

We enter now the topic of the precision measurement of
$\theta_{23}$. According to the discussion of section \ref{sec2}, we expect
a variation of the rate of electron neutrino events (and in particular of
SubGeV electron neutrinos) as a function of $\theta_{23}$. In the simple
case of vacuum oscillations and $\theta_{13}$ we expect: $\Delta N_e \propto (1 -
N^0_\mu/N^0_e sin^2 \theta_{23})$, where $N^0_\mu$ and $N^0_e$ are the
unoscillated $\mu$-like and $e$-like event rates. 
The situation is slightly more complex
for $\theta_{13}$ = 0 and when matter effects are present. Our complete
simulation gives the behavior shown in Fig. \ref{nu5}, where the deviation
from no-oscillation prediction of event rate of SubGeV $e$-like events (in
events/kton yr) is plotted vs $sin^2\theta_{23}$.

\begin{figure}
\begin{center}
\includegraphics[width=10cm]{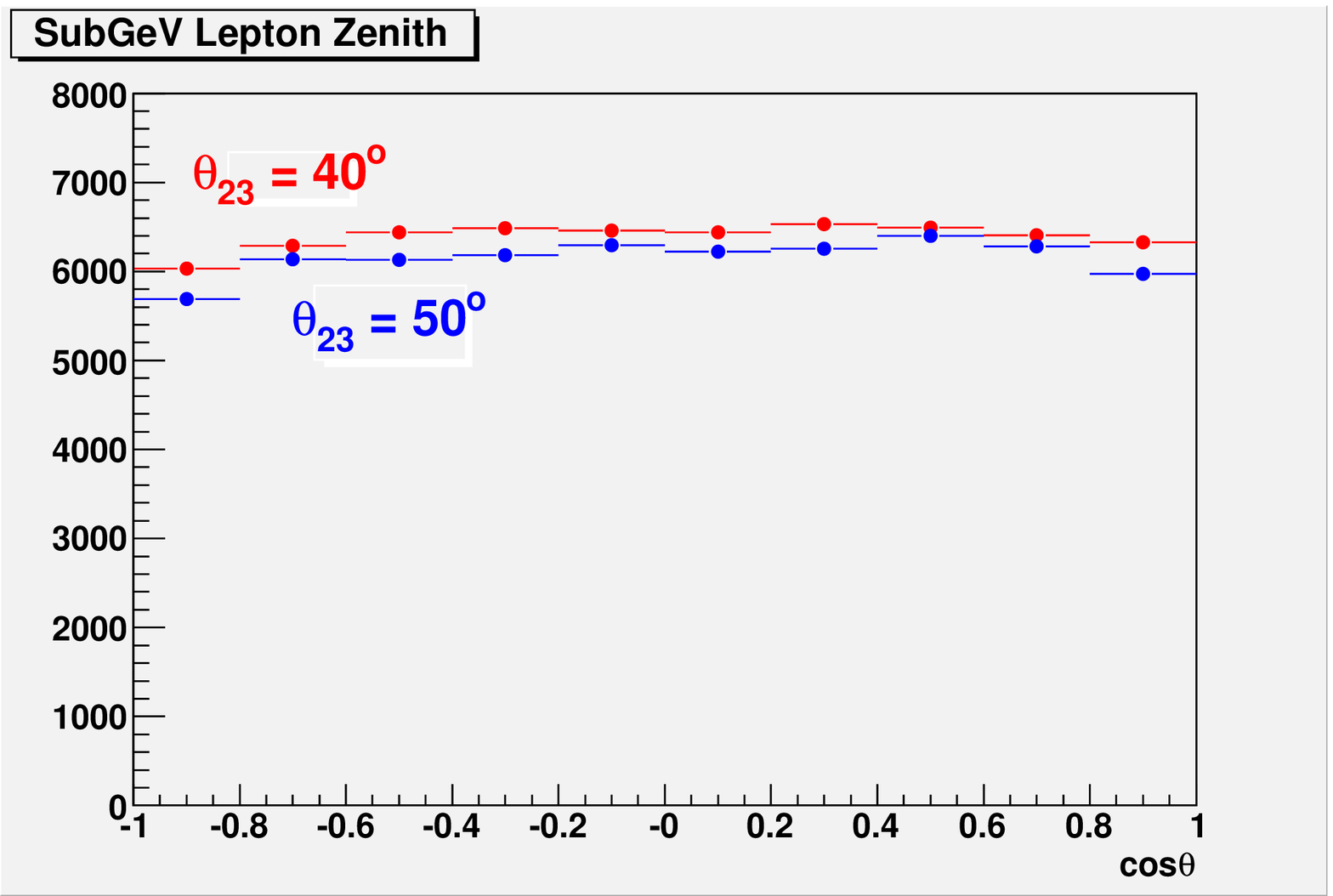}
\caption{Absolute prediction of the zenith angular distribution of SubGeV
  $e$-like events for two different values of $\theta_{23}$ when
  $\theta_{13}$ = 0$^\circ$.
\label{nu6}}
\end{center}
\end{figure}

As expected, the relative differences in rate are very small and a very
large exposure is needed. The absolute difference in rate of SubGeV
$e$-like events is shown in Fig. \ref{nu6}, for $\theta_{13}$~=~0 and for two
values of $\theta_{23}$ symmetric with respect to 45$^\circ$ (only the
lepton direction is considered here).

\begin{figure}
\begin{center}
\includegraphics[width=10cm]{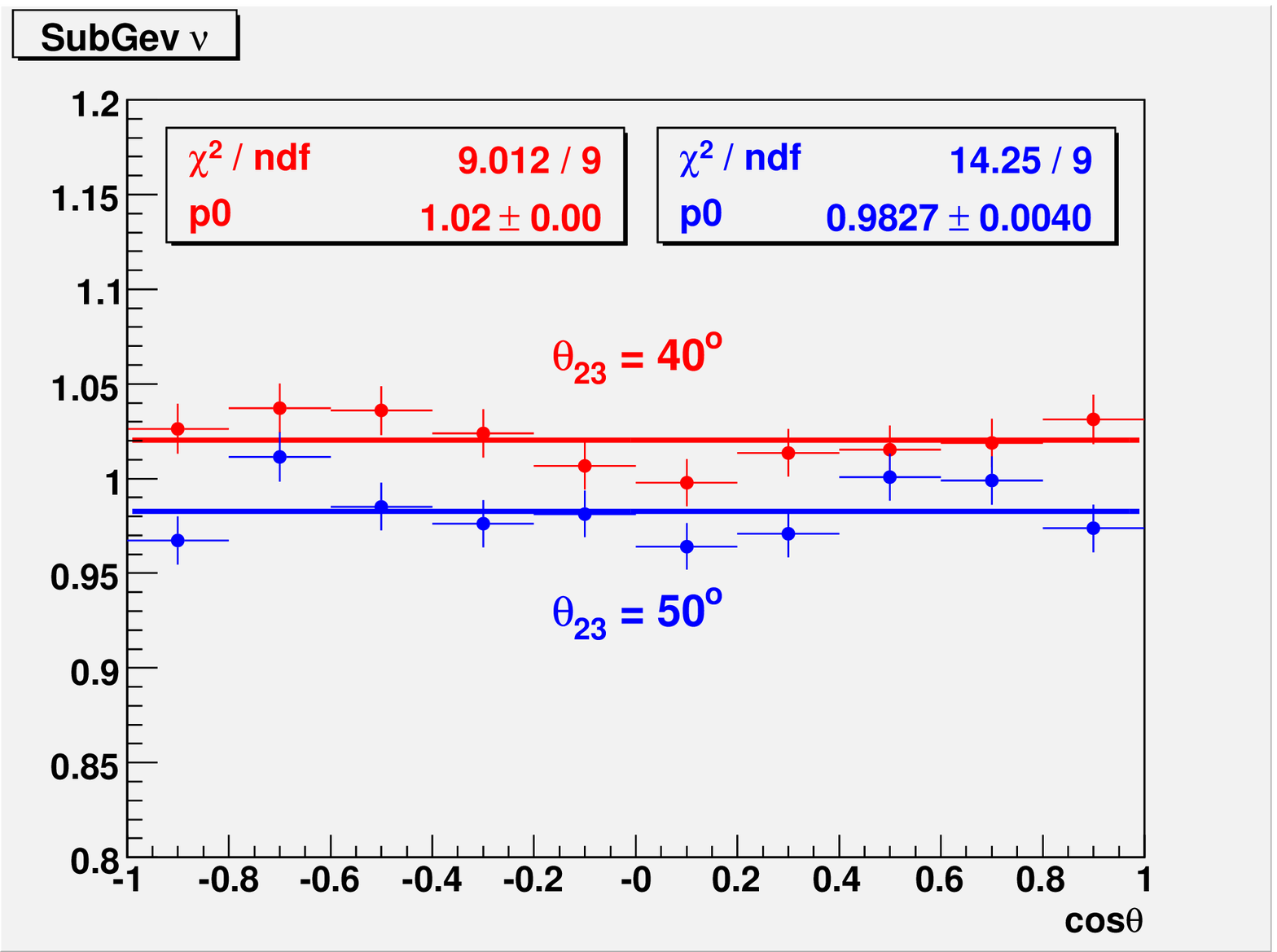}
\caption{$N_e/N^0_e$ ratio as a function of the zenith angle of SubGeV
  $e$-like events for two different values of $\theta_{23}$ when
  $\theta_{13}$ = 0$^\circ$.
\label{nu7}}
\end{center}
\end{figure}

If we take the ratio with respect to the non oscillated prediction, we get
the result plotted in Fig.\ref{nu7}

\begin{figure}
\begin{center}
\begin{tabular}{c}
\includegraphics[width=10cm]{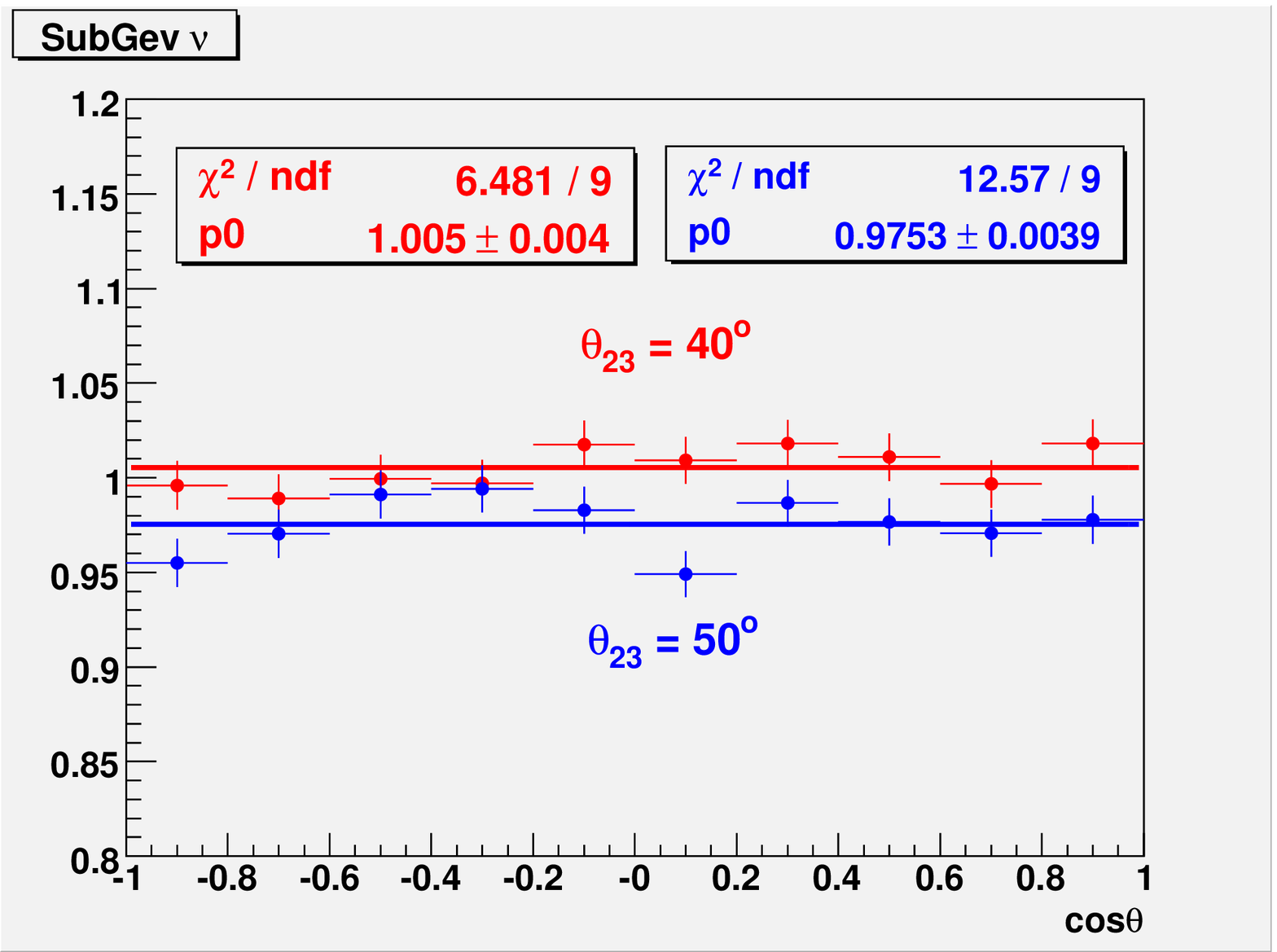} \\
\includegraphics[width=10cm]{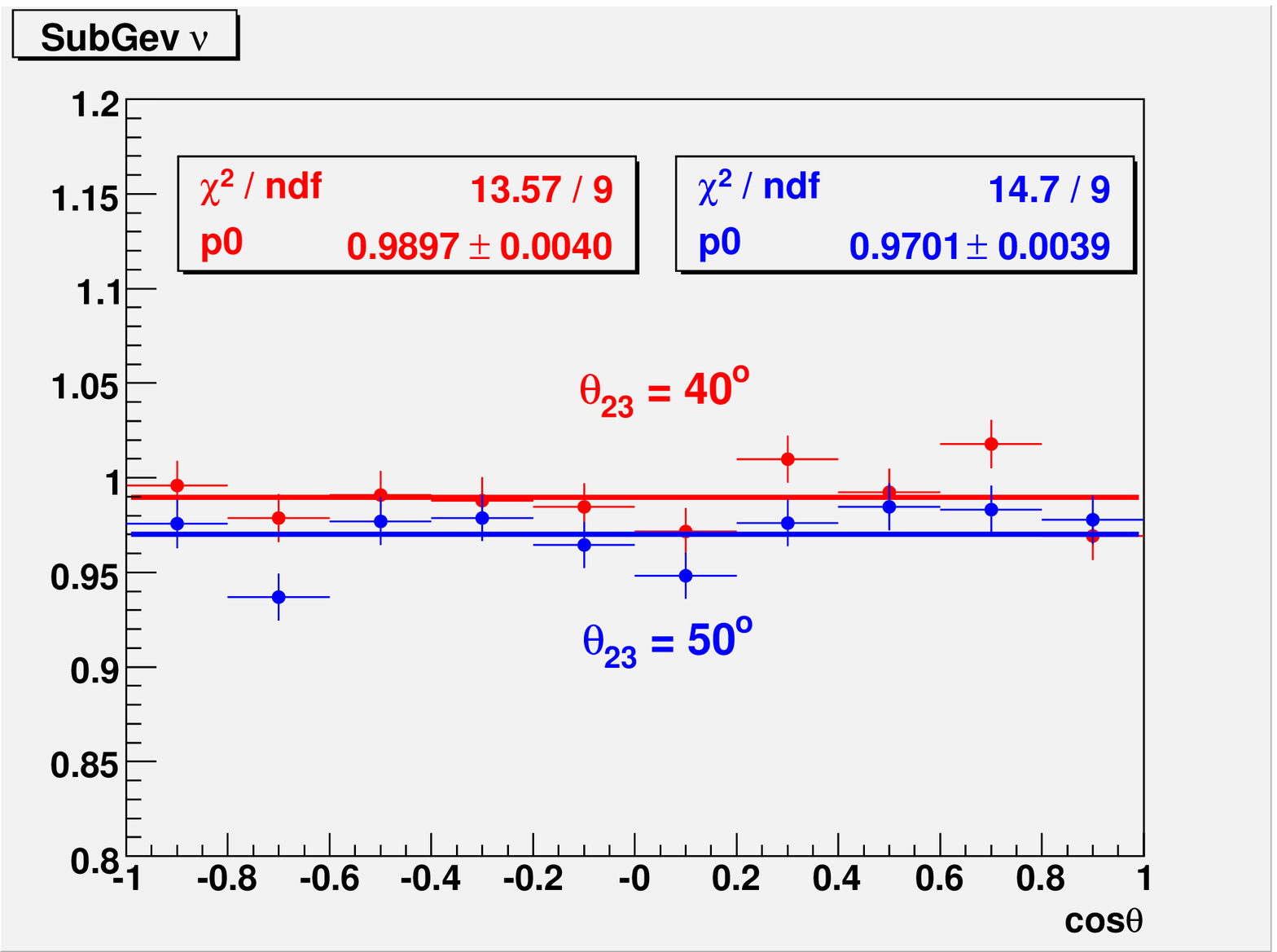} 
\end{tabular}
\caption{$N_e/N^0_e$ ratio as a function of the zenith angle of SubGeV
  $e$-like events for two different values of $\theta_{23}$ when
  $\theta_{13}$ = 5$^\circ$ (top) and $\theta_{13}$ = 10$^\circ$.
\label{nu8}}
\end{center}
\end{figure}

If we conservatively fit the SubGeV distribution with a constant (namely
 this is equivalent to 
 integrating the angular distribution over the whole range of the zenith
 angle) 
the difference in
 the ratio $N_e/N^0_e$ between $\theta_{23}$ = 40$^\circ$ and 50$^\circ$ is
 at the level of 0.037 
$\pm$ 0.006. For relatively large values of $\theta_{13}$, the significance
 in the difference becomes smaller, as shown in Fig.\ref{nu8}.

\begin{figure}
\begin{center}
\includegraphics[width=10cm]{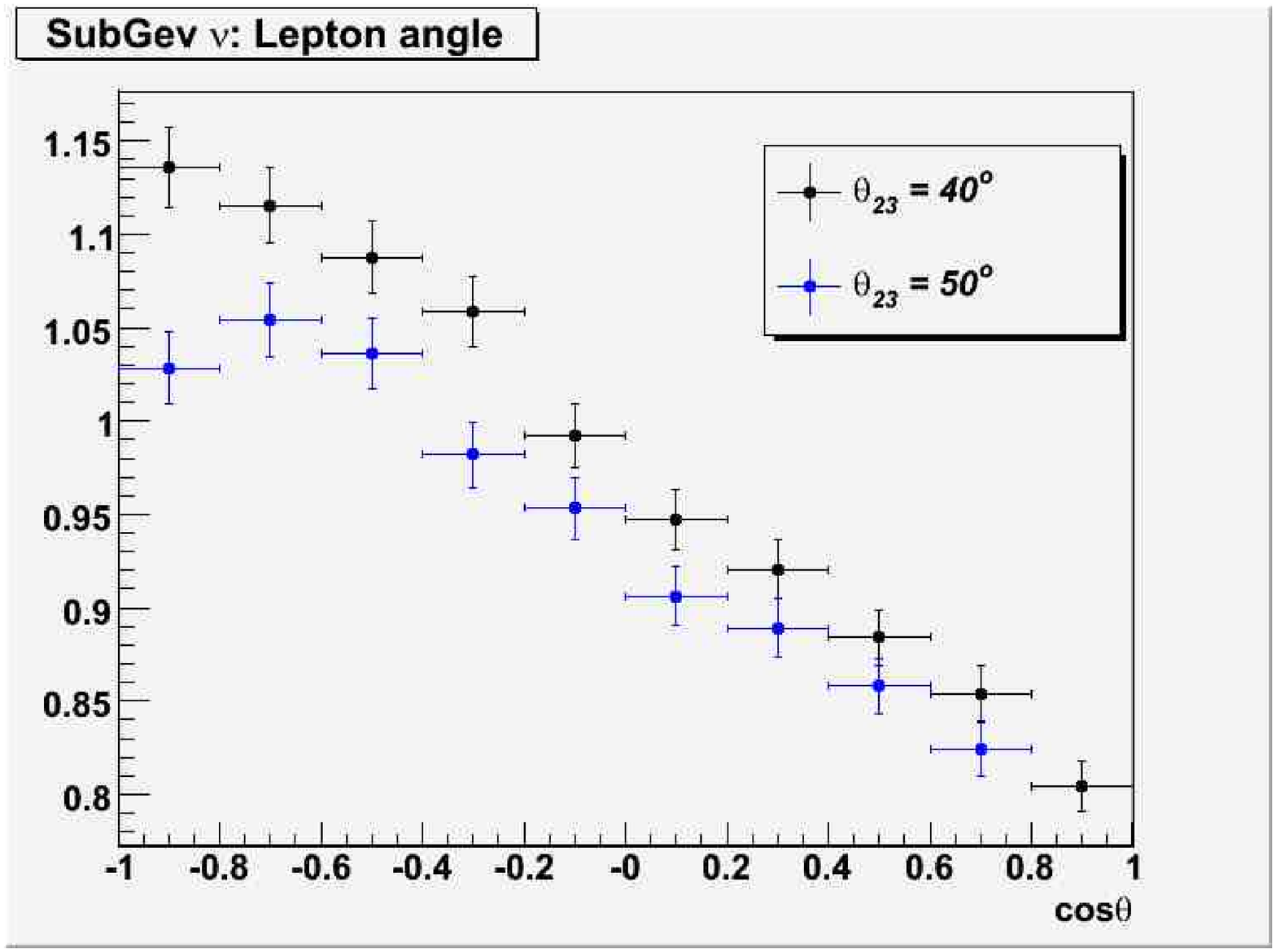}
\caption{Double ratio $(N_e/N^0_e)/(N_\mu/N^0_\mu)$ as a function of the
  zenith angle of SubGeV $e$-like events for two different values of
  $\theta_{23}$ when $\theta_{13}$ = 0$^\circ$.
\label{nu9}}
\end{center}
\end{figure}

Of course it is hard to believe that one could rely on the absolute level of $N_e$ 
prediction: flux normalization remains one of the most important 
uncertainties. A better analysis is obtained by the double ratio:
$(N_e/N_\mu)/(N^0_e/N^0_\mu)$. The variation of $\mu$-like events has the
opposite trend of $N_e$, and in first approximation all common
uncertainties contributing both to $N_e$ and $N_\mu$ cancel out. Here the
most important uncertainties concern the 
normalization of the primary cosmic ray flux and the knowledge of
$\nu$-Nucleus cross sections. In good approximation these affect $e$ and
$\mu$ flavor almost at the same level (there are mass dependent terms which
are of course non negligible at very low energy).
As an example of the possible expectation,
 we get the behavior as
a function of the zenith angle which shown in Fig.\ref{nu9}
(when $\theta_{13}=0^\circ$).

Now the important topic, as far as systematic uncertainties are concerned,
is the accuracy on the prediction of the $N_e/N_\mu$ ratio. 
This topic has been already debated in the context of a dedicated workshop
stimulated by Super--Kamiokande\cite{rcc04}. One of the reported
conclusions was that the measurement of $\theta_{23}$ 
octant can be done achieving an exposure of at least 20 years of runs of
Super--Kamiokande in order to distinguish (at the level of
$\Delta\chi^2\sim$2) between the two mirror values corresponding to $\sin^2
2\theta_{23}$=0.96 ($\sim$39$^\circ \div$ 51$^\circ$), with the present
level of systematics. 
An effort to improve the accuracy on the prediction on $N_\mu$/$N_e$ at a
level better than 1\% would be very helpful.

\section{Summary and discussion}
\label{sec6}
A very large LAr TPC allows to detect low energy neutrinos (above $\sim$ 50
MeV) with null or negligible
experimental systematic error. Therefore, in principle, such a detector
can give new important contributions to
neutrino  physics, also by means of a new investigation of atmospheric neutrinos.
An exposure of at least 500 kton yr is however necessary in order to 
give new significant contributions. The sector of SubGeV $\nu_e$, which is
particularly suitable for an ICARUS--like detector, offers in particular 
the possibility of performing new precision  measurements.
In this work we have performed a preliminary investigation about the
possibility of the measuring how close is $\theta_{23}$ to the full mixing
value of 45$^\circ$. In principle this measurement is affordable if two
conditions are met: an exposure larger or equal to 500 kton yr and a
reduction of systematic uncertainties about neutrino fluxes, where the most 
important item is the knowledge of the $N_e/N_\mu$ ratio.

\end{document}